\documentclass[%
 reprint,
nofootinbib,
 amsmath,amssymb,
 aps,
]{revtex4-2}

\usepackage{psfrag,color,epsfig}
\usepackage{dcolumn}
\usepackage{graphicx,graphics}	% Including figure files
\usepackage{amsmath}	% Advanced maths commands
\usepackage{amssymb}	% Extra maths symbols
\usepackage{bm}		% Bold maths symbols, including upright Greek
\usepackage{multirow}
\usepackage{tabularx}
\usepackage{tikz}
\usetikzlibrary{matrix,calc}
\usepackage{physics}
\usepackage{xspace}
\usepackage{soul} % strike out
\usepackage[T1]{fontenc}
\usepackage{ae,aecompl}

\usepackage{newtxmath}
\usepackage{ulem}
\makeatletter
\DeclareRobustCommand{\Hi}{%
  \mbox{H\check@mathfonts\fontsize\sf@size\z@\selectfont I}%
}

\def\kk{{\bm{k}}}
\def\qq{{\bm{w}}}
\def\mm{{\bm{u}}}
\def\nn{{\bm{v}}}

\def\x{\bm{x}}
\def\X{\bm{x}_{\perp}}

\def\mpc{{\rm Mpc}}

\def\Nt{N_{\rm tri}}
\def\pp{\hat{\bm{p}}}

\long\def\/*#1*/{}

\begin{document}
% \label{firstpage}
% \pagerange{\pageref{firstpage}--\pageref{lastpage}}

%--------------------------------------------------------------------
\preprint{APS/123-QED}
\title[Fast estimator of redshift space bispectrum multipoles ]{Brisk estimator for the angular multipoles of the redshift space bispectrum}

%-------------------------------------------------------

\author{Sukhdeep Singh Gill}
 \altaffiliation[]{sukhdeepsingh5ab@gmail.com}%Lines break automatically or can be forced with \\
\author{Somnath Bharadwaj}%
 \email{somnathbharadwaj@gmail.com}
\affiliation{%
 Department of Physics, Indian Institute of Technology Kharagpur, Kharagpur 721302, India.
}%

% \author{Charlie Author}
%  \homepage{http://www.Second.institution.edu/~Charlie.Author}
% \affiliation{
%  Second institution and/or address\\
%  This line break forced% with \\
% }%
% \affiliation{
%  Third institution, the second for Charlie Author
% }%

%-----------------------------------------
\begin{abstract}
The anisotropy of the redshift space bispectrum depends upon the orientation of the triangles formed by three $\kk$ modes with respect to the line of sight. For a triangle of fixed size ($k_1$) and shape ($\mu,t$), this orientation dependence can be quantified in terms of angular multipoles $B_\ell^m(k_1,\mu,t)$ which contain a wealth of cosmological information. We propose a fast and efficient FFT-based estimator that computes the bispectrum multipole moments $B_\ell^m$ of a 3D cosmological field for all possible $\ell$ and $m$ (including $m\neq 0$). The time required by the estimator to compute all multipoles from a gridded data cube of volume $N_g^3$ scales as $\sim \mathcal{O}(N_g^4)$ in contrast to the direct computation technique which requires time $\sim \mathcal{O}(N_g^6)$. Here, we demonstrate the formalism and validate the estimator using a simulated non-Gaussian field for which the analytical expressions for all the bispectrum multipoles are known. The estimated results are found to be in good agreement with the analytical predictions for all $16$ non-zero multipoles (up to $\ell= 6, m=6$). We expect the $m \neq 0$ bispectrum multipoles to significantly enhance the information available from galaxy redshift surveys and future redshifted 21-cm observations. 
\end{abstract}
%--------------------------------------------------------------------
\maketitle
%--------------------------------------------------------------------
\date{\today}%
%-------------------------------------------
\section{Introduction }
\label{sec:intro}

The spatial distribution of galaxies has emerged as one of the primary probes of large-scale structure (LSS) of the Universe. The simplest model of inflation predicts the primordial density fluctuations that seed the LSS to be a Gaussian random field. Nevertheless, the subsequent non-linear growth and non-linear biasing induce non-Gaussianity in the galaxy distribution. Further, many inflationary paradigms speculate non-Gaussian primordial fluctuations. In any scenario, it is essential to exploit the statistics that capture the non-Gaussian nature of galaxy distribution and utilize the maximum potential of data from surveys such as SDSS\footnote{https://www.sdss.org/} \cite{2000AJYork}, DESI\footnote{https://www.desi.lbl.gov/} \cite{Levi_2013}, LSST\footnote{https://www.lsst.org/}   \cite{Ivezic_2019}, and EUCLID\footnote{https://www.euclid-ec.org/} \cite{Laureijs_2011}. The bispectrum is the lowest-order statistic sensitive to non-Gaussianity in the underlying distribution. It is a function of the closed triangles formed by three $\kk$ vectors and contains the information on mode coupling. The inclusion of the bispectrum significantly supplements our comprehension of the standard model of cosmology, enabling robust constraints on the primordial non-Gaussianity \citep{Sefusatti_2009,Fergusson_2012,Oppizzi_2018, Planck_NG_2018,Shiraishi_2019,Feldman_2001, Scoccimarro_2004, Liguori_2010, Scoccimarro_2015, Ballardini_2019, Pearson_2018}, modeling the inflation \citep{2004_Babich,2004_Bartolo,2023_Sohn_DHOST}, lifting the degeneracy between $\Omega_m$ and $b_1$ \citep{Scoccimarro_1999}, constraining cosmological parameters \citep{ivanov_2023,2022_Philcox_oneloop,2022_Guido_oneloop}, and galaxy bias parameters \citep{Matarrese_1997}. 

The observed galaxy bispectrum is anisotropic along the line of sight (LoS) direction because of the peculiar motion of galaxies. This anisotropy of the redshift space bispectrum can be studied by decomposing it into spherical harmonics $Y_\ell^m(\hat{p})$  \citep{1999_Scoccimarro,2020_Bharadwaj,2020_Mazumdar}. The expansion coefficients in $Y_\ell^m(\hat{p})$ basis, the bispectrum multipole moments $B_\ell^m$, can be measured from the survey data. While the majority of work has focused on the monopole moment $B_0^0$ only (e.g. \citep{2022_Philcox_oneloop}), a few have considered the higher order multipoles
\citep{Byun_2022,Tellarini_2016,ivanov_2023,2022_Guido_oneloop}. 
\citet{ivanov_2023} have shown that including the $\ell=2$ and $4$ multipoles 
 in the analysis only leads to a marginal improvement of the cosmological parameters.  However, the studies mentioned above have all been restricted to the $m=0$ multipoles. \citet{Gagrani_2017} has studied the information loss associated with restricting the study to $m=0$ multipoles using the Fisher information formalism and find that $m\neq0$ multipoles do not carry significant information. 
 However, in a recent work, \citet{Heinrich_2023} has shown that incorporating the $m\neq 0$ multipole moments significantly improves the information gain. It is crucial to accurately measure the higher-order multipoles with $m\neq 0$ for precision cosmology.  \citet{Nan_2018} have used the halo model to make analytical predictions for all the non-zero higher-order multipoles  (including those with $m\neq 0$) up to $\ell=4$, $m=4$,  
and compared these with the values computed from mock galaxy samples. 
In this paper, we present a fast and accurate estimator to calculate the bispectrum multipoles $B_\ell^m$, including those with $m \neq 0$,  
considering triangles of all possible unique shapes.  To the best of our knowledge, this is the first implementation of an FFT-based fast bispectrum estimator that includes the  $m\neq 0$ multipoles. This paper builds on the formalism and bispectrum parameterization presented in \citep{2020_Bharadwaj} and \citep{2020_Mazumdar}.

A brief outline of the paper is as follows. Section~\ref{sec:meth} presents the Methodology where we describe the formalism and algorithm of the estimator. Section~\ref{sec:validation} presents the Validation of the estimator, including theoretical background, data simulation, and binning effects. Section~\ref{sec:summary} presents the Summary and Discussion.

%%%%%%%%%%%%%%%%%%%%%%%%%%%%%%%%%%%%%%%%%%%%%%%%%%%%%%%%%%%%%%%%%%%%%%%%%%%%%%%%

\section{Methodology }
\label{sec:meth}
%\subsection{Formulation}

The bispectrum $B^s(\kk_1,\kk_2,\kk_3)$ of any random field in redshift space $\delta^s(\x)$ distributed in a box of volume $V$ is defined as,
\begin{equation}
 B^s(\kk_1,\kk_2,\kk_3)=V^{-1} \langle \Delta^s(\kk_1) \Delta^s(\kk_2) \Delta^s(\kk_3) \rangle,
\label{eq:bs}
\end{equation}
with $\kk_1+\kk_2+\kk_3=0$ imposing that three $\kk$ vectors form a closed triangle.  $\langle \cdots \rangle$ denotes the average with respect to an ensemble of independent realizations, and $\Delta^s(\kk)$ is the Fourier transform of $\delta^s(\x)$. Decomposing $\kk \equiv  (\kk_\perp, k_\parallel)$   into components $\kk_\perp$ and $k_\parallel$ that are respectively perpendicular and parallel to the LoS direction $\hat{z}$, $\Delta^s(\kk_\perp,k_{\parallel})$ is statistically isotropic only on the plane perpendicular to $\hat{z}$. 
The closed triangle condition  yields  two conditions  $k_{1\parallel}+k_{2\parallel}+k_{3\parallel}=0$ and $\kk_{1\perp}+\kk_{2\perp}+\kk_{3\perp}=0$. The first condition implies that 
we can independently fix only two $k_{\parallel}$, namely  $k_{1\parallel}$ and $k_{2\parallel}$, for which the third one is determined using $k_{3\parallel}=-k_{1\parallel}-k_{2\parallel}$. The second condition implies that the perpendicular components of the three wave vectors must form a closed triangle. The bispectrum, which  is invariant under a rotation along $\hat{z}$, depends only on the magnitudes of  $\kk_{1\perp},\kk_{2\perp}$ and $\kk_{3\perp}$ , namely $k_{1 \perp},k_{2 \perp}$ and $k_{3 \perp}$. The bispectrum is thus a function of 5 independent parameters, and we may express it as $B^s(k_{1 \perp},k_{2 \perp},k_{3 \perp},k_{1\parallel},k_{2\parallel})$. 

Here we opt for the parameterization prescribed by \cite{2020_Bharadwaj}. For a triangle with $k_1\geq k_2\geq k_3$, its size is parameterized by $k_1=|\kk_1|$, shape by two dimensionless parameters 
\begin{equation}
\mu =- \frac{\kk_1 \cdot \kk_2}{k_1 k_2}, \hspace{0.5cm} \, t = \frac{k_2}{k_1} \,,
\label{eq:shape}
\end{equation}
and orientation with respect to $\hat{z}$   by a unit vector $\pp$. The cosine of the angle between the three sides of the triangle
 and  $\hat{z}$ are respectively given by 
\begin{eqnarray}
&\mu_{1} & = p_z \,,  \hspace{0.5 cm} \, 
\mu_{2} = -\mu p_z + \sqrt{1-\mu^2}p_x \nonumber \\
&\mu_{3}& = \dfrac{-[(1-t\mu)p_z+t\sqrt{1-\mu^2}p_x]}{\sqrt{1-2t\mu+t^2}}
\label{eq:orient}
\end{eqnarray}

Theoretically, we calculate the expansion coefficients of the bispectrum in terms of $Y_\ell^m(\pp)$ the spherical harmonic  basis using  \citep{2020_Bharadwaj}
\begin{equation}
    \bar{B}_\ell^m(k_1,\mu,t )=\sqrt{\dfrac{2\ell+1}{4\pi}}{\int[Y_\ell^m(\hat{\textbf{p}})]^*~B^s(\hat{\textbf{p}},k_1,\mu,t )~d\Omega_{\hat{\textbf{p}}}},
\label{eq:bs_lm}
\end{equation}
where the integration is over $4\pi$ steradian, corresponding to all possible orientations of a triangle of a fixed size $(k_1)$ and shape $(\mu,t)$. The integration assumes the continuum limit for the $\kk$ modes. Hereafter,  we refer to the results obtained by analytically performing this integral as $[\bar{B}_\ell^m]_{\rm c}$. From the observational point of view, we consider a finite volume $V$ with $N_g^3$ grid points.  
The multipole moments can be calculated by identifying the set $\mathcal{T}$ of all triangles of a fixed size $(k_1)$ and shape $(\mu,t)$, and using 
\begin{equation}
[\bar{B}_\ell^m]_{\rm d}(k_1, \mu, t)=\sqrt{\frac{2\ell+1}{4\pi}}\frac{\sum_{\mathcal{T}} [Y_\ell^m(\pp)]^* B^s(\pp,k_1, \mu, t)}{\sum_{\mathcal{T}} |Y_\ell^m(\pp)|^2}~,
\label{eq:bs_lm2}
\end{equation}
where $\sum_{\mathcal{T}}$ denotes a sum over all triangle orientations that are accessible in the 3D gridded data cube. The multipole moments $[\bar{B}_\ell^m]_{\rm c}$ (Eq.~\ref{eq:bs_lm}) do not  exactly match  $[\bar{B}_\ell^m]_{\rm d}$ (Eq.~\ref{eq:bs_lm2}) because of the discrete sampling of $\kk$ space whereby the 
 integration is replaced with a summation \citep{ivanov_2023,2022_Ivanov_preciseRSDBS}. We have accounted for this effect and will describe it in the next section.

\begin{figure*}
    \centering 
    \includegraphics[width=.98\textwidth]{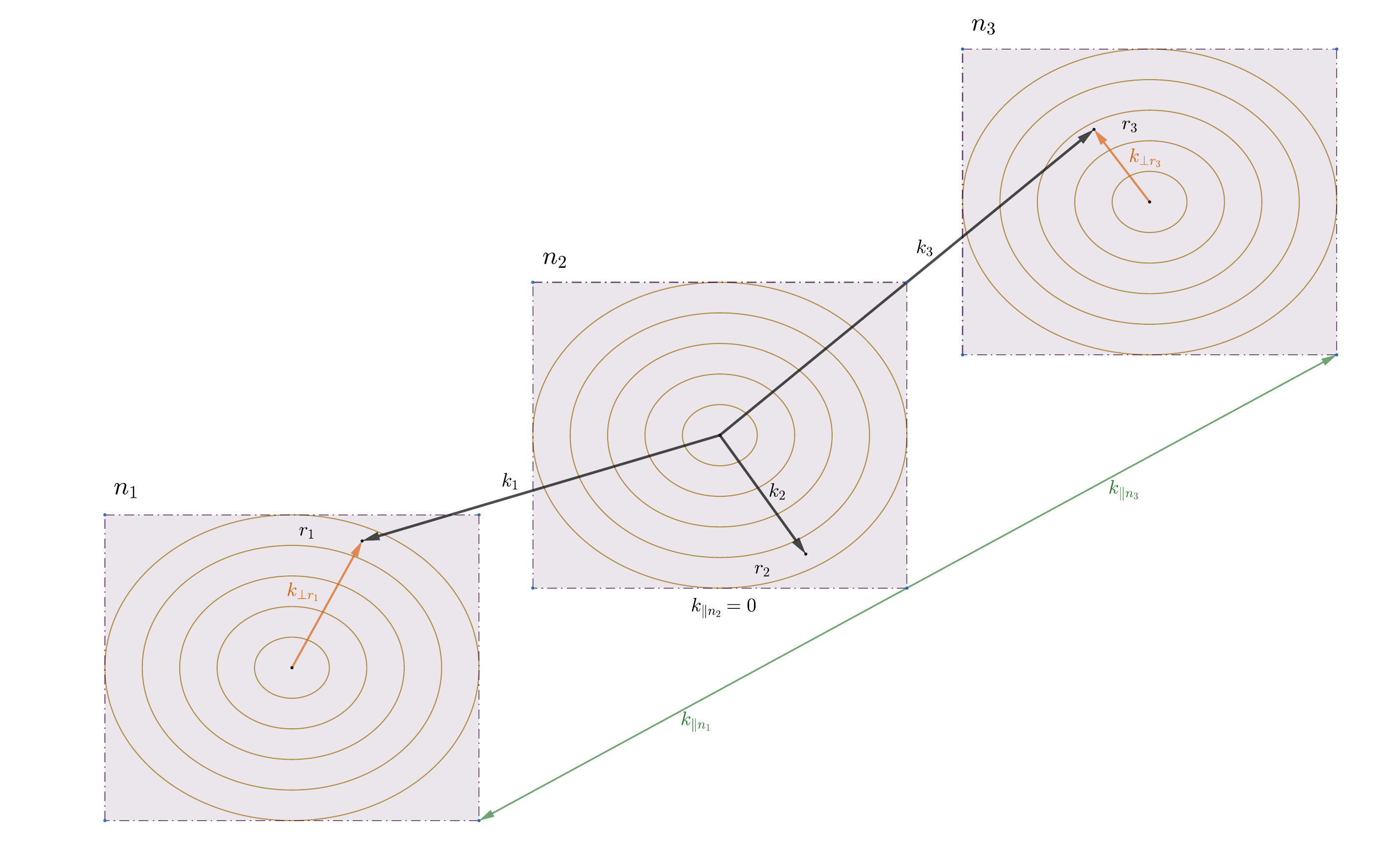} 
    \caption{The binning scheme of the estimator. This demonstrates the division of $\kk$ space into multiple planes perpendicular to the Line of Sight (LoS) direction $\hat{z}$, with each plane further subdivided into concentric annular rings. We show a closed triangle formed by three modes $\kk_1,\kk_2,\kk_3$ that fall in different annular rings labeled as $r_1,r_2,r_3$. The combination of three rings ($r_1,r_2,r_3$) and planes ($n_1,n_2,n_3$) defines a bin of triangles that have a fixed shape ($\mu,t$), size ($k_1$), and orientation ($\pp$) with respect to the LoS. If this triangle is rotated around the LoS such that three $\kk$ modes are restricted within respective annular rings, then the rotated triangle will still have the same orientation ($\pp$) relative to the LoS as that of the original triangle. The bin contains all such triangles. The $\kk$ vectors are decomposed into their $k_{\parallel}\hat{z}$ and $\kk_\perp$ components with respect to the LoS. Note that the middle plane is considered to be centered at the origin $\kk=0$ (hence $k_{\parallel n_2}=0$) to avoid complexity in the illustration.}
    \label{fig:binning_scheme}
\end{figure*}
	
%and take $\hat{\bm{z}}$ as the LoS direction. In principle, to calculate various multipoles corresponding to a triangle of a given shape and size, we have to identify all such triangles formed by $\kk$ modes in the box and use Eq.\ref{eq:bs_lm2}. However,
The direct implementation of Eq.~\ref{eq:bs_lm2} is immensely time-consuming as it involves identifying triangles, which requires computations of order  $N_g^6$. Several earlier works by \cite{Sefusatti_thesis,Jeong_thesis,Scoccimarro_2015} have proposed an FFT based technique to compute binned statistics in Fourier space, which considerably speeds up the bispectrum computation. Following a similar formalism, \citep{2021_Shaw} has proposed a fast estimator that considers spherical shells in $\kk$ space to compute 
the binned bispectrum monopole $\bar{B}_0^0(k_1, \mu, t)$.  Further, this can be easily extended to estimate the higher multipoles with $m=0$, as has been implemented in \citep{2024_gill}. However,  the  $m\neq 0$ multipoles introduce additional complexity due to $e^{i m\phi}$ phase factor in the spherical harmonics, which encapsulates the dependence on the azimuthal direction. It is not possible to incorporate this phase factor if we use spherical shells in $\kk$ space to compute the binned bispectrum. We have overcome this difficulty by considering a different binning scheme, as shown in Fig.~\ref{fig:binning_scheme}. Here, the box can be viewed as a collection of $N_g$ 2D planes normal to $\hat{z}$, which we label using $n=1,2,3,...,N_g$ each with a corresponding $k_{\parallel n}$. Each plane is %uniformly 
divided into $N_r$ concentric annular rings, which we label using $r=1,2,3,...,N_r$. Considering the annular ring $r$,  $\{r\}$ denotes the set of $\kk_{\perp}$ modes contained in the ring, $k_{\perp r}$ denotes its mean radius and $(\delta k_\perp)_r$ its width.  The largest ring is restricted to a maximum outer radius of $N_g/3$  by the periodic boundary condition imposed on the finite volume  \citep{2021_Shaw}. 
The annular rings and three planes are schematically shown in Fig.~\ref{fig:binning_scheme}. 

 To estimate the binned bispectrum, we consider three annular rings  $( r_1, r_2, r_3)$  respectively located in the planes $(n_1,n_2,n_3)$.  As discussed earlier, only $n_1$ and $n_2$  are independently specified, $n_3$ is determined from the condition $k_{\parallel n_1}+k_{\parallel n_2}+k_{\parallel n_3}=0$. 
We define the binned bispectrum estimator as
\begin{equation}
\begin{aligned}
& \Hat{B}^s(k_{\perp r_1},k_{\perp r_2},k_{\perp r_3},k_{\parallel n_1},k_{\parallel n_2}) 
= \frac{1}{V~\Nt} \sum_{\mm }     \sum_{\nn} \times  \\&  \sum_{\qq} 
\Delta(\mm,k_{\parallel n_1})  \Delta(\nn,k_{\parallel n_2}) 
 \Delta(\qq,k_{\parallel n_3}) \,  \delta_{\rm K}(\mm+\nn+\qq) ,
\end{aligned}
\label{eq:binbs}
\end{equation}
where  $\mm \in \{r_1\}, \nn \in \{r_2\}$ and $ \qq \in \{r_3 \}$  refer to three 2D wave vectors  in the plane perpendicular to $\hat{z}$.  The 2D Kronecker delta function $\delta_{\rm K}(\mm+\nn+\qq)$ ensures that the estimator receives a contribution only when the three vectors $(\mm,\nn,\qq)$ form a closed triangle and $\Nt$ is the total number of such triangles. 
The binned bispectrum estimator  (eq.~\ref{eq:binbs}) provides an estimate of the bispectrum averaged over a set of triangles that have nearly the same shape, size, and orientation with respect to $\hat{z}$.

The FFT implementation of Eq.~\ref{eq:binbs} utilizes the Fourier expansion of the 2D 
Kronecker delta function
\begin{equation}
\begin{aligned}
    \delta_{\rm K}(\mm+\nn+\qq)
    &= \frac{1}{N_{\rm g}^2} \sum_{\X} \exp(-i [\mm+\nn+\qq] \cdot \X)~.
    \label{eq:kdelta}
\end{aligned}
\end{equation}

It is now possible to use this to express the estimator as 
\begin{equation}
\begin{aligned}
 & \Hat{B}^s(k_{\perp r_1},k_{\perp r_2},k_{\perp r_3},k_{\parallel n_1},k_{\parallel n_2}) 
= \frac{1}{V~\Nt}\frac{1}{N_{\rm g}^2} \times \\
&\quad 
\sum_{\X} D(r_1,n_1,\X)  D(r_2,n_2,\X)  D(r_3,n_3,\X)  ~,
\end{aligned}
\label{eq:fftbs}
\end{equation}
where 
\begin{equation}
    D(r_1,n_1,\X) =\sum_{\mm \in \{r_1\}} \Delta(\mm,k_{\parallel n_1})   \exp(-i\mm \cdot\X)~, 
    \label{eq:iFT}
\end{equation} 
which is the 2D inverse FFT of the field $\Delta(\mm,k_{\parallel n_1})$ with $\mm$ restricted to the annular ring $r_1$.  The other $D$s are similarly defined.

% \begin{equation} 
% 	M(\kk_{a_n\perp}) =\begin{cases}
% 1 &   \text{if }  k_\perp \text{ lie within ring } a_n \\
% 0 &  \text{otherwise}
% 	\end{cases}
% 	\end{equation}

We now calculate  $\Nt$, the total number of triangles  that appears in Eq.~(\ref{eq:fftbs}), using  
\begin{equation}
\Nt = \sum_{\mm\in \{r_1\}}  \sum_{\nn\in \{r_2\}} \sum_{\qq \in \{r_3\}} \delta_{\rm K}(\mm+\nn+\qq)~.
\label{eq:Ntri1}
\end{equation}
We have implemented Eq.~(\ref{eq:Ntri1}) using 
\begin{equation}
    \Nt =\frac{1}{N_{\rm g}^2} \sum_{\X} I(r_1,\X)  I(r_2,\X)  I(r_3,\X)  ~,
\label{eq:Ntri2}
\end{equation}
where 
\begin{equation}
     I(r_1,\X) =\sum_{\mm \in \{r_1\}}   \exp(-i\mm \cdot\X)~, 
    \label{eq:2FT}
\end{equation} 
and the other $I$s are similarly defined.

 The particular bin under consideration, for which we have calculated the binned bispectrum  $\Hat{B}^s(k_{\perp r_1},k_{\perp r_2},k_{\perp r_3},k_{\parallel n_1},k_{\parallel n_2})$    considers all closed triangles where the three sides, namely  $(\mm,k_{\parallel n_1})$, $(\nn,k_{\parallel n_2})$ and $(\qq,k_{\parallel n_3})$, have $(k_{\parallel n_1},k_{\parallel n_2},k_{\parallel n_3})$ fixed, whereas  $(\mm,\nn,\qq)$   lie within the annular rings  $(r_1,r_2,r_3)$  respectively.  All these triangles have nearly the same shape and size. 
Although the triangles formed by  $(\mm,\nn,\qq)$  have different orientations in the plane perpendicular to $\hat{z}$, they all have nearly similar values of $(\mu_1,\mu_2,\mu_3)$, or equivalently $\pp$   (Eq.~\ref{eq:orient}),  which decides the inclination with respect to  $\hat{z}$.  We use Eq.~\ref{eq:shape} and Eq.~\ref{eq:orient} to transform from $(k_{\perp r_1},k_{\perp r_2},k_{\perp r_3},k_{\parallel n_1},k_{\parallel n_2})$ to $(\hat{\textbf{p}},k_1,\mu,t )$, whereby we have the binned bispectrum estimator $ \hat{B}^s(\pp,k_1, \mu, t)$ that we use in Eq.~\ref{eq:bs_lm2} to calculate the multipole moments $\hat{\bar{B}}_\ell^m(k_1, \mu, t)$.

%\subsection{Implementation} \label{sec:implementation}
%In this subsection, 
We now briefly discuss the implementation of the bispectrum estimator.
The definition of the annular rings is an important factor that affects the accuracy of the estimator, as well as its computational complexity. Here, we have chosen the ring width $(\delta k_\perp)_r$, in grid units,  to start from $1$ and double after every $10$ rings
 i.e., $(\delta k_\perp)_r=2^{\lfloor(r-1)/10\rfloor}$, where $\lfloor. \rfloor$ denotes the floor function.  As mentioned earlier, the largest ring is restricted to a maximum outer radius of $N_g/3$  by the periodic boundary condition imposed on the finite volume  \citep{2021_Shaw}. In this setting, $N_r$ has a  logarithmic dependence on $N_g$  i.e. $N_r \sim \log(N_g)$. 
 
 We have implemented the estimator in two steps. In Step I, we compute the function $D(r,n,\x_\perp)$ (Eq.~\ref{eq:iFT}) and store it for each ring on all the planes. The computation time to compute a $D(r,n,\x_\perp)$, which is a 2D FFT, scales as $\sim N_g^2 \, \log(N_g)$. Considering the $N_r$ rings and the $N_g$ planes involved in Step I, 
 the computation time scales as $T_{\rm I} \sim N_r \times N_g \times N_g^2 \, \log(N_g)$. However, $N_r$ has a weak logarithmic dependence on $N_g$. Ignoring the logarithmic dependence, we have  $T_{\rm  I} \sim N_g^3$. We also compute and store $I(r,\x_\perp)$  (Eq.~\ref{eq:2FT}) that does not vary with $n$, and it thus suffices to evaluate this for a single plane. The computation time for this scales as $ N_g^2$, which is subdominant, and we ignore this contribution to $T_{\rm I}$.   
 
 In Step II, we consider all possible combinations of three rings and two planes.  We implement Eq.~\ref{eq:fftbs} to compute  $ \hat{B}^s(\pp,k_1, \mu, t)$ for each distinct combination of rings and planes. The number of possible combinations scales as $N_r^3 \times N_g^2$, and the sum $\sum_{\X}$ scales as $N_g^2$. Combining these factors, and ignoring the logarithmic dependence, we have $T_{\rm II} \sim N_g^4$. 
 
 The values of $(k_1, \mu, t)$ at which we have the estimates of the binned bispectrum  $ \hat{B}^s(\pp,k_1, \mu, t)$  are not uniformly distributed. We have implemented a second level of binning to combine these to estimate the angular multipoles ( Eq.~\ref{eq:bs_lm2}).  
 We have divided the $k_1$ range into bins of equal logarithmic interval and also divided the range $0.5 \le \mu,t \le 1$ into bins of equal linear interval.  
 The exact number of bins that we have used here varies with $\ell$, and it has been adjusted to achieve a high signal-to-noise ratio without smearing out the shape and size dependence of the signal. Considering Eq.~\ref{eq:bs_lm2}, we have implemented the second level of binning using 
 \begin{equation}
\hat{\bar{B}}_\ell^m(\bar{k}_1, \bar{\mu},\bar{t}) =\sqrt{\frac{2\ell+1}{4\pi}}
\frac{\sum [Y_\ell^m(\pp)]^*\hat{B}^s(\pp,k_1, \mu, t) \Nt} {\sum |Y_\ell^m(\pp)|^2 \Nt}
\label{eq:bs_lm2_1}
\end{equation}
 where $(\bar{k}_1, \bar{\mu},\bar{t})$ in  the l.h.s. refers to a particular interval in $(k_1, \mu, t)$ respectively, and $\hat{\bar{B}}_\ell^m$ refers to the bin average estimate evaluated with respect to this interval. The sum in the r.h.s. refers to the estimates $\hat{B}^s(\pp,k_1, \mu, t)$ that occur in that particular interval. Note that 
 $\Nt$ (Eq.~\ref{eq:Ntri2}) depends on $(\pp,k_1, \mu, t)$, however, for the brevity of notation, we have not shown this explicitly. 

 In our implementation, we do not store the values of $ \hat{B}^s(\pp,k_1, \mu, t)$  as the data volume is large.   For each estimate of the binned bispectrum  $ \hat{B}^s(\pp,k_1, \mu, t)$, we identify the appropriate $(\bar{k}_1, \bar{\mu},\bar{t})$   bin, and  directly accumulate both the numerator and the denominator of Eq.~\ref{eq:bs_lm2_1}
 for the required values of $\ell$ and $m$.

\begin{figure}
    \centering 
    \includegraphics[width=.47\textwidth]{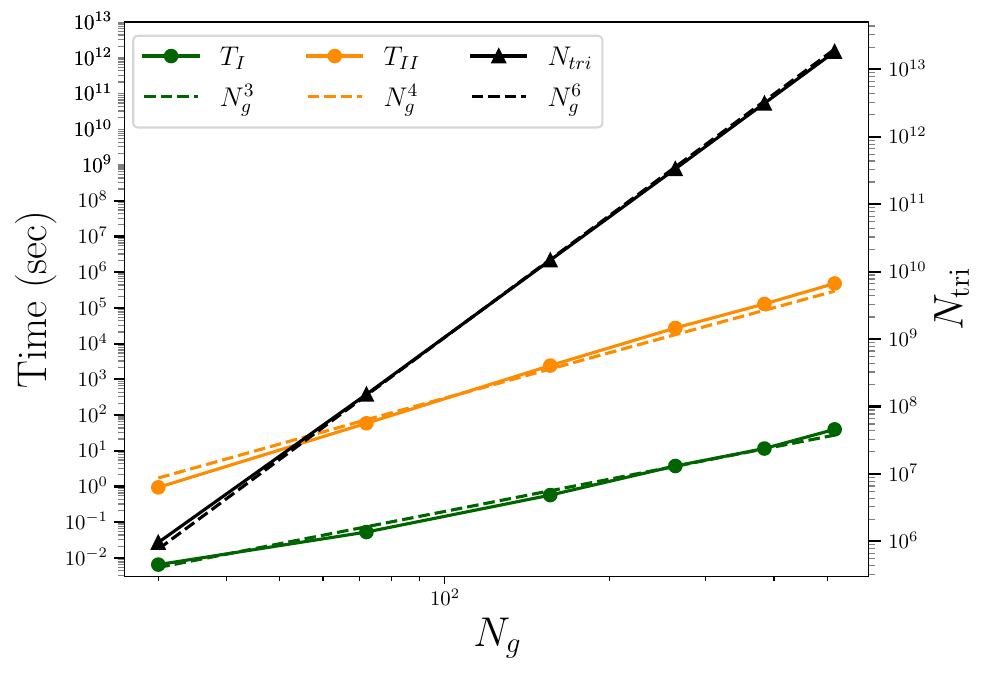} 
    \caption{ Computation time for the bispectrum multipoles as a function of the number of grids $N_g$. The solid green line shows $T_I$ the time taken to perform all the FFT operations (Step I), and the solid orange line shows $T_{II}$ the time to compute the bispectrum multipoles (Step II) considering all possible closed triangles. The corresponding dashed lines show the expected scaling with $N_g$, as indicated in the figure legend. The solid black line shows the total number of the closed triangles $N_{\rm tri}$, and the dashed black line shows $ N_g^6$. The computation has been  performed on a single-core CPU.} 
    \label{fig:runtime}
\end{figure}

We next discuss the computational complexity of the estimator. The computation time required for a direct estimate of the bispectrum multipoles by summing over $\kk_1$ and $\kk_2$ and identifying each closed triangle is expected to scale as $\sim N_{\rm tri}\propto N_g^6$. The solid black line in Fig.~\ref{fig:runtime} shows $N_{\rm tri}$, the total number of triangles computed using Eq.~\ref{eq:Ntri2}, as a function of $N_g$.  We see that this closely matches the expected $N_{\rm tri}\propto N_g^6$ behavior shown by the dashed black line.   The solid green line shows $T_I$ the time taken to perform all 2D FFT operations in Eq.~\ref{eq:iFT} and Eq.~\ref{eq:2FT} and save them to disk. We see that this matches the expected $T_I \propto N_g^3$ behavior 
shown by the dashed green line. The solid orange line shows $T_{II}$ the time taken to compute Eq.~\ref{eq:fftbs} and Eq.~\ref{eq:Ntri2} for all possible combinations of rings and planes, and subsequently use these to compute all the non-zero bispectrum multipole moments up to $\ell=6$ and $m=6$.  We see that this matched the $T_{II} \propto N_g^4$ line shown by the dashed orange line.  
The slight deviation between the solid and dashed lines most probably arises from the  $\log(N_g)$ terms that have not been considered when considering the scaling behavior of the computation time. 

 We see (Fig.~\ref{fig:runtime}) that the total computation time is dominated by Step II for which $T_{II} \propto N_g^4$. 
The results for the computation time presented in this work have been performed on a single-core CPU. However, both Step I and Step II can be easily parallelized. Considering Step I,  the 2D FFT operations in Eq.~\ref{eq:iFT} and Eq.~\ref{eq:2FT} for the different rings and planes can each be executed in parallel on a different thread. Taking into account Step II, Eq.~\ref{eq:fftbs} and Eq.~\ref{eq:Ntri2} for all possible combinations of three rings and planes, each combination can be executed in parallel on a different thread. Ideally, we expect the computation time to be reduced approximately by the number of cores for a parallel implementation of the estimator. However, oftentimes the
performance per core drops when using many concurrently, and the actual gain may be less than expected.

%, thereby significantly reducing the overall computation time.
% In our fast estimator (Eq.~\ref{eq:fftbs}), we perform 2D Fourier transforms (Eq.~\ref{eq:iFT}) for all rings on each plane. The computation time to perform a 2D FFT scale as $\sim N_g^2\log(N_g)$ \citep{2021_Shaw}, and considering $N_g$ planes,  the total time to perform the FFTs scales as $\sim N_g^3\log(N_g)$. Note that $N_r \ll N_g$, and the

\section{Validating the estimator}
\label{sec:validation}
\subsection{Theoretical expressions}
To validate the estimator, we generate a mildly non-Gaussian random field in redshift space $\delta^s(\x)$ for which the bispectrum multipole moments are analytically derived. We start with generating a Gaussian random field in real space $\delta^r_{\rm G}(\x)$ using a  specified power spectrum $P(k)$ and introduce non-Gaussianity through a local quadratic bias, 
\begin{equation}
    \delta^r(\x)=\delta^r_{\rm G}(\x)+f_{\rm NG}  \big\{[\delta^{r}_{\rm G}(\x)]^2-\langle [\delta^{r}_{\rm G}(\x)  ]^2\rangle\big\}~
\label{eq:nG}
\end{equation}
where the parameter $f_{\rm NG}$ controls the level of non-Gaussianity in the random field $\delta^r(\x)$. The real space bispectrum $B^r$ computed,  to linear order in $f_{\rm NG}$, is
\begin{equation}
\begin{aligned}
B^r&(k_1,k_2,k_3)\\&= 2 f_{\rm NG} [P(k_1)P(k_2)+P(k_2)P(k_3)+P(k_3)P(k_1)]~.
\end{aligned}
\label{eq:mod_bs}
\end{equation}

The field is mapped to redshift space by applying the linear RSD,
\begin{equation}
    \Delta^s(\kk)=(1+\beta_1 \mu_\kk^2)\Delta^r(\kk)~. 
    \label{eq:RSD}
\end{equation}
where $\Delta^r(\kk)$ and $\Delta^s(\kk)$ are the Fourier transform of the field in real and redshift space, respectively, and  $\beta_1$ is the linear RSD parameter. The linear RSD theory predicts the redshift space bispectrum $B^s(\hat{\textbf{p}},k_1,\mu,t )$ for this field to be,
\begin{equation}
\begin{aligned}
B^s&(\hat{\textbf{p}},k_1,\mu,t )\\&=(1+\beta_1 \mu_{1}^2)(1+\beta_1 \mu_{2}^2)(1+\beta_1 \mu_{3}^2)B^r(k_1,k_2,k_3) ~.
\label{eq:rsd_bs}
\end{aligned}
\end{equation}
% where $\mu_i=(\hat{z}\cdot \kk_i)/|\kk_i|$ is the cosine of the angle between $\hat{z}$ and wave vector $\kk_i$ with $i = 1, 2, 3$. 

The analytical expressions for the multipole moments $B_\ell^m(k_1,\mu,t)$ can be calculated using Eq.~(\ref{eq:bs_lm}). These expressions are rather lengthy and, hence, not explicitly shown here. We refer the reader to Eqs. (24-29,31-33,A1-A7) of \cite{2020_Bharadwaj} for details. The analytical expressions are expected to hold true for sufficiently small values of $f_{\rm NG}$ such that $f_{\rm NG} \, \sigma_{\rm G} \ll 1$, where ${\sigma^2_{\rm G} = \langle [\delta^r_{\rm G}]^2 \rangle}$ is the variance of ${\delta^r_{\rm G}}(\x)$. 
% \subsection{Cosmic Variance}
% Cosmic variance can be written as,
% \begin{equation}
%     \sigma^2_{B} =\frac{1}{\Nt}[{V P(k_1)P(k_2)P(k_3) + 3 B_{Ana}^2(k_1,k_2,k_3)}]~,
% \label{eq:cv}
% \end{equation}

\subsection{Simulations}
We simulated the field in a box of volume  $V=L^3=(215~\mpc)^3$ divided in $N_{\rm g}=384$ grids in each direction, for which we have a grid spacing of   $0.56~\mpc$, and $k_{\rm min}=0.029~\mpc^{-1}$.  We use the input power spectrum $P(k)=k^{-2}$ to generate $\delta_{\rm G}^r$. The non-Gaussianity parameter is chosen to be $f_{\rm NG}=0.5$ for which $f_{\rm NG} \, \sigma_{\rm G} \approx 0.2$, and set the value of the linear RSD parameter $\beta_1=1$. We have generated $50$ independent realizations of the field and used these to estimate the mean   $\bar{B}_\ell^m(\bar{k}_1, \bar{\mu},\bar{t}) = \langle \hat{\bar{B}}_\ell^m(\bar{k}_1, \bar{\mu},\bar{t}) \rangle $ and also the   $1 \sigma$ error bars. 

\subsection{Binning and discreteness effects}
\label{sec:bin_disc}
In this work, we have validated the estimator by comparing the values of  
$\hat{\bar{B}}_\ell^m(\bar{k}_1, \bar{\mu},\bar{t})$
%$\bar{B}_\ell^m(k_1, \mu, t)$ 
estimated from the simulations with the analytical predictions for the same quantities. However, the analytical predictions 
$[\bar{B}_\ell^m]_{\rm c}$  presented in \citep{2020_Bharadwaj}, which have been obtained by using  Eq.~(\ref{eq:rsd_bs}) in  Eq.~(\ref{eq:bs_lm}), assume the continuum limit.  In reality, we only have discrete $\kk$ modes, and the integral in Eq.~(\ref{eq:bs_lm}) is, in principle, replaced by the sum in  Eq.~(\ref{eq:bs_lm2}). In the present work, we consider the binned bispectrum where the binning takes place in two steps. In the first step, we compute the binned bispectrum 
$\hat{B}^s(k_{\perp r_1},k_{\perp r_2},k_{\perp r_3},k_{\parallel n_1},k_{\parallel n_2})$  (Eq.~\ref{eq:binbs}). Here Eq.~(\ref{eq:fftbs}) directly yields the bispectrum averaged over a bin of triangles; these triangles all have nearly the same shape, size, and orientation with respect to $\hat{z}$. In the second step, we transform the variables
$ (k_{\perp r_1},k_{\perp r_2},k_{\perp r_3},k_{\parallel n_1},k_{\parallel n_2}) \rightarrow (\hat{\textbf{p}},k_1,\mu,t )$ and use Eq.~(\ref{eq:bs_lm2_1}) to calculate $\hat{\bar{B}}_\ell^m(\bar{k}_1, \bar{\mu},\bar{t})$  
where $(\bar{k}_1, \bar{\mu},\bar{t})$ refers to a particular interval in $(k_1, \mu, t)$ respectively, and $\hat{\bar{B}}_\ell^m$ refers to the average evaluated with respect to this interval. This average, which has $[Y_\ell^m(\pp)]^* \Nt$ as weights, combines triangles with nearly the same shape and size, but with different orientation with respect to $\hat{z}$. 

To account for the effect of binning, we introduce the discrete analytical predictions $[\bar{B}_\ell^m]_{\rm d}(\bar{k}_1, \bar{\mu},\bar{t})$ (as denoted by the subscript ${\rm d}$), which have been calculated by 
using Eq.~(\ref{eq:mod_bs}) in  Eq.~(\ref{eq:rsd_bs}). The resulting predictions have been binned using Eq.~(\ref{eq:bs_lm2_1}) in exactly the same way as for the estimator.  Note that  $[\bar{B}_\ell^m]_{\rm d}(\bar{k}_1, \bar{\mu},\bar{t})$  only incorporates the binning in the second step and $[\bar{B}_\ell^m]_{\rm d}(\bar{k}_1, \bar{\mu},\bar{t})$ does not incorporate the binning in the first step (Eq.~\ref{eq:fftbs}). This would require us to perform the $\sim N_g^6$ computation that explicitly calculates the binned bispectrum (Eq.~\ref{eq:binbs}).

We validate the estimator by comparing $\bar{B}_\ell^m(k_1, \mu, t)$ with the discrete analytical predictions $[\bar{B}_\ell^m]_{\rm d}(k_1, \mu, t)$.
Note that here and in the subsequent discussion,
for brevity of notation,  we have used $(k_1, \mu, t)$
instead of $(\bar{k}_1, \bar{\mu},\bar{t})$ as arguments for the final binned bispectrum.

Most theories which model the bispectrum typically provide analytical predictions for $[\bar{B}_\ell^m]_{\rm c}$.  Therefore, it is often desirable to compare the estimated values $\bar{B}_\ell^m(k_1,\mu,t)$ with the analytical predictions $[\bar{B}_\ell^m]_{\rm c}$ and not $[\bar{B}_\ell^m]_{\rm d}$.
 A few earlier  works  \citep{2022_Ivanov_preciseRSDBS,ivanov_2023} have addressed this issue 
by introducing  `discreteness weights' $w_\ell^m (k_1,\mu,t)$ 
\begin{equation}
    w_\ell^m(k_1,\mu,t)=\dfrac{[\bar{B}_\ell^m]_{\rm d}(k_1,\mu,t)
    }{[\bar{B}_\ell^m]_{\rm c}(k_1,\mu,t)}
\end{equation}
which depend on the multipole moment $(\ell,m)$, the triangle size and shape $(k_1,\mu,t)$ and possibly, also the theory under consideration. We have not considered this here, and we have directly compared $\bar{B}_\ell^m$ with  $[\bar{B}_\ell^m]_{\rm d}$.

\subsection{Results}

\begin{figure}
    \centering 
    \includegraphics[width=.47\textwidth]{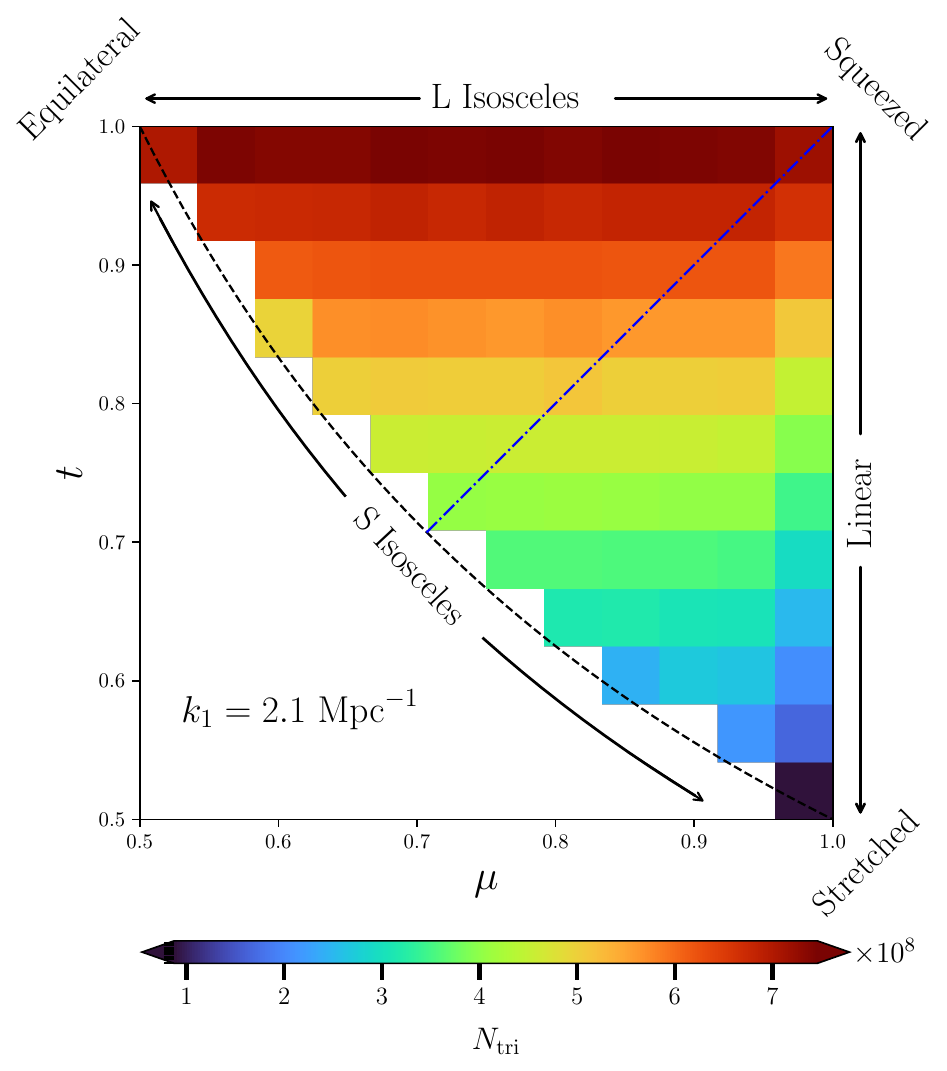} 
    \caption{Shows the total number of triangles $\Nt$ of all possible shapes $(\mu,t)$ with a fixed size $k_1=2.1~ {\rm Mpc}^{-1}$.}
    \label{fig:Ntri2}
\end{figure}
 The $[\bar{B}_\ell^m]_{\rm c}$ values are predicted to have a scale-invariant power-law $k_1$ dependence, and it is thus adequate to show the results for just a single value of $k_1$. The number of triangles scales as $\Nt \propto k_1^6$ \citep{2023_Mazumdar}, and we have chosen a large value $k_1=2.1~ {\rm Mpc}^{-1}$   with a bin width  $dk_1=0.2~{\rm Mpc}^{-1}$ so as to achieve a large value of $\Nt$ and obtain a reliable (high SNR) estimate of $\bar{B}_\ell^m$. So as to accurately capture the shape dependence of 
$\bar{B}_\ell^m(k_1,\mu,t)$, we have divided  the $(\mu,t)$ plane into fine bins of width $(d\mu,dt)\approx (0.042,0.042)$ for evaluating the $\ell =0$ and $2$ multipoles. The values of $\bar{B}_\ell^m$ and the SNR are smaller for the higher multipoles $\ell >2$,  and we have used coarser $(\mu,t)$ bins for these.

 Fig.~\ref{fig:Ntri2} illustrates the $(\mu,t)$ bins that we have used to parameterize the triangle shape dependence of $\bar{B}_\ell^m(k_1,\mu,t)$   
 for $\ell \le 2$. It also shows $\Nt$ the number of triangles corresponding to each bin. Considering the $\mu-t$ plane, the $\mu=1$ boundary corresponds to linear triangles where the two larger sides of the triangle are nearly aligned in the same direction. The $t=1$ and $2\mu t=1$ borders correspond to the L  and S Isosceles triangles, respectively. The two largest sides are equal for L Isosceles triangles, whereas the two smaller sides are equal for S. 
  We see that in all cases, the number of triangles is of the order of $10^8$, and it is largest for L Isosceles triangles. Further,  $\Nt$ decreases almost monotonically with the decline in $t$, and the number falls by a factor of $10$ as triangles are deformed to the stretched limit. Nonetheless, $\Nt$ does not vary significantly with $\mu$.

% \begin{figure*}
%     \centering 
%     \includegraphics[width=.9\textwidth]{fig/fft_ana_mut_l=0.pdf} 
%     \caption{Estimator binning scheme }
%     \label{fig:mu_t_l=0}
% \end{figure*}

% \begin{figure*}
%     \centering 
%     \includegraphics[width=.9\textwidth]{fig/fft_ana_mut_l=2.pdf} 
%     \caption{Estimator binning scheme }
%     \label{fig:mu_t_l=2}
% \end{figure*}

% \begin{figure*}
%     \centering 
%     \includegraphics[width=.8\textwidth]{fig/fft_ana_mut_l=4_5X5.pdf} 
%     \caption{Estimator binning scheme }
%     \label{fig:mu_t_l=4}
% \end{figure*}

% \begin{figure*}
%     \centering 
%     \includegraphics[width=.6\textwidth]{fig/fft_ana_mut_l=6_5X5.pdf} 
%     \caption{Estimator binning scheme }
%     \label{fig:mu_t_l=6}
% \end{figure*}

% \begin{figure*}
%     \centering 
%     \includegraphics[width=1\textwidth]{fig/ps_mom.pdf} 
%     \caption{ }
%     \label{fig:ps_mom}
% \end{figure*}

\begin{figure*}
    \centering 
    \includegraphics[width=1\textwidth]{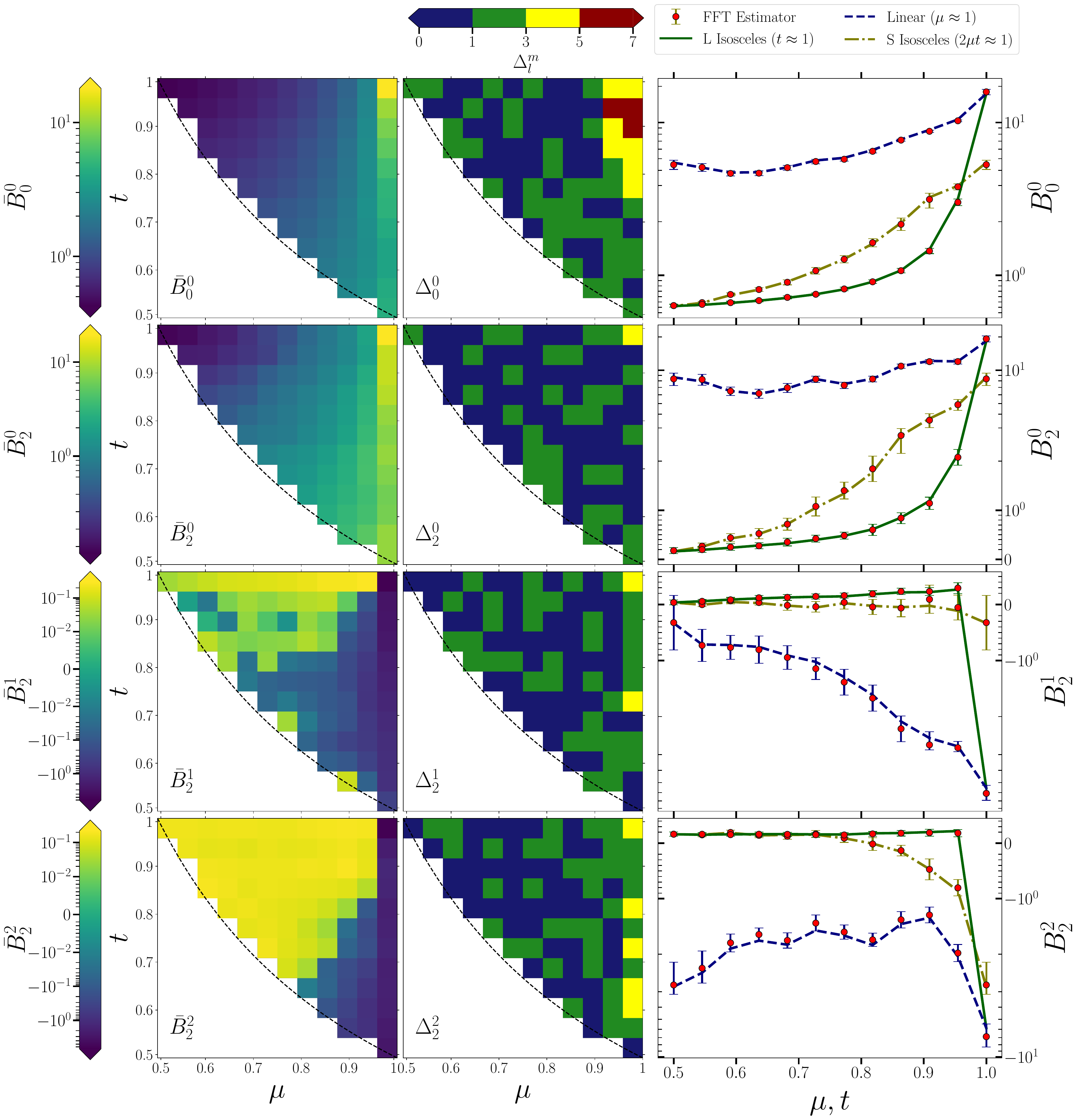} %fft_ana_mut_l0_2_v3_rev.pdf
    \caption{ Results for $\ell=0$ and $2$ multipoles. It shows the estimates and their deviation from analytical prediction for all triangle shapes ($\mu,t$) having a fixed size $k_1=2.1~ {\rm Mpc}^{-1}$. The uppermost row depicts the results for the monopole ($\ell=0,m=0$). The second, third, and fourth rows show $\ell=2$ and $m=0,1,2$ moments, respectively. The first column illustrates the mean estimates $\bar{B}_\ell^m$ from simulations. The second column shows $\Delta_\ell^m=|\Delta {B}_\ell^m|/{\sigma} _{B_\ell^m}$, $\Delta {B}_\ell^m$ is the deviation of the estimates from analytical predictions and ${\sigma}_{B_\ell^m}$ is the expected rms  statistical fluctuation. The third column explicitly shows the estimated ${B}_\ell^m$ values (Red points with  $5{\sigma}_{B_\ell^m}$ error bars) 
    for linear, L, and S Isosceles triangles.
    The various lines show the corresponding analytical predictions.  Results are plotted against $\mu$ for L and S Isosceles triangles and against $t$ for linear triangles.}
    \label{fig:mu_t_l=0&2}
\end{figure*}

 Fig.~\ref{fig:mu_t_l=0&2} shows the results for the $\ell=0$ and $2$ multipoles. Here, each row corresponds to a different combination of $\ell$ and $m$, as indicated in the figure.  
 The panels in the left column show  $\bar{B}_{\ell}^m$ estimated from the simulations. The analytical predictions  $[\bar{B}_\ell^m]_{\rm d}$ are visually indistinguishable from the $\bar{B}_{\ell}^m$ shown here, and we have not explicitly shown 
 $[\bar{B}_\ell^m]_{\rm d}$. The panels in the middle column   show
 $\Delta_\ell^m=| \Delta {B}_\ell^m |/\sigma_{B_\ell^m}$ where 
$\Delta \bar{B}_\ell^m={\bar{B}}_\ell^m - [\bar{B}_\ell^m]_{\rm d}$ is the difference between the estimated value and the analytical predictions, and $\sigma_{B_\ell^m}$ (estimated from the $50$ realizations of the simulations) is the rms  statistical fluctuations expected in
$\Delta \bar{B}_\ell^m$.  $\Delta_\ell^m$ provides an estimate of whether the deviations between the estimator and analytical predictions are consistent with the expected statistical fluctuations or not. The panels in the right column show a comparison of $\bar{B}_{\ell}^m$ and 
$[\bar{B}_\ell^m]_{\rm d}$ along the three boundaries of the $\mu-t$ plane. 

We first consider the first row of
Fig.~\ref{fig:mu_t_l=0&2}, which shows the results for $\ell=0,m=0$. We see that the value of the estimated 
$\bar{B}_{0}^0$ is minimum for equilateral triangles  $ ((\mu,t)=(0.5,1))$, and it increases 
as $\mu$ increases towards linear triangles $(\mu=1)$. The top-most curve of the right panel shows 
the analytical prediction $[\bar{B}_0^0]_{\rm d}(t)$ along the right boundary $(\mu \approx 1)$ of the  $\mu-t$ plane. We see that  $\bar{B}_{0}^0$ has the largest values along this curve. The value of  $[\bar{B}_0^0]_{\rm d}(t)$  further increases as we move from stretched triangles $((\mu,t)=(1,0.5))$ to squeezed triangles $((\mu,t)=(1,1))$ along this curve.  The bottom-most curve shows $[\bar{B}_0^0]_{\rm d}(\mu)$
 along the top boundary $(t \approx 1)$ of the  $\mu-t$ plane. Proceeding from equilateral to squeezed triangles along this curve, we see that $[\bar{B}_0^0]_{\rm d}$
  increases gradually until $\mu \approx 0.9$, beyond which it increases sharply close to the squeezed limit. The middle curve shows $[\bar{B}_0^0]_{\rm d}(\mu)$
  along the bottom   boundary $(2 \mu t \approx  1)$ of the  $\mu-t$ plane. We see that the value of $[\bar{B}_0^0]_{\rm d}$  increases gradually as we proceed from equilateral to stretched triangles along this curve.  In all cases, the  estimated values $\bar{B}_{\ell}^m$ 
  coincide with the  analytical predictions $[\bar{B}_\ell^m]_{\rm d}$, and the fractional deviations are  within the range $[-0.033, 0.033]$. Considering the middle panel, we find that $\Delta^0_{0} \le 3$ across most of the $\mu-t$ plane, which indicates that the differences between the estimated values and the analytical predictions are largely within the expected statistical fluctuations. We have 
  a few bins where  $3 < \Delta^0_{0} \le 5$, which also may be consistent with the expected statistical fluctuations. In addition, there are a few bins where 
   $5 < \Delta^0_{0} \le 7$ where the deviations are in excess of those expected from statistical fluctuations. The $3$ bins where $5 < \Delta^0_{0} \le 7$ are all located near the squeezed limit, and a similar behavior has been reported in earlier work \citep{2021_Shaw}. 
   The reason for this behavior is that the bispectrum is very sensitive to the triangle shape near the squeezed limit. For a particular combination of rings $(r_1,r_2,r_3)$ and planes $(n_1,n_2,n_3)$, the binned estimator (Eq.~\ref{eq:fftbs}) includes triangles of different shapes due to the finite width of the annular rings  (Fig.~\ref{fig:binning_scheme}). In contrast, as discussed in Section~\ref{sec:bin_disc}, 
   the discrete analytic prediction $[\bar{B}_\ell^m]_{\rm d}$ 
   only considers the bispectrum at the mean value of $(k_{\perp r_1},k_{\perp r_2},k_{\perp r_3},k_{\parallel n_1},k_{\parallel n_2})$ corresponding to the combination of three annular rings and planes. Note, however, that $[\bar{B}_\ell^m]_{\rm d}$ incorporates the second level of binning (Eq.~\ref{eq:bs_lm2_1}) that is used to calculate  $\hat{\bar{B}}_\ell^m$.

We next consider the quadrupole moment ($\ell=2$)  with $m=0$, which is shown in the second row of  Fig.~\ref{fig:mu_t_l=0&2}.  We see that the values of 
$\bar{B}_{2}^0$  are comparable to those of $\bar{B}_0^0$. 
The results for $\bar{B}_{2}^0$ are also very similar to those for $\bar{B}_{0}^0$, and the right and left panels of the second row are very similar to the corresponding panels of the first row.  Considering the middle panel, we see that the deviations from the analytical predictions are within the expected statistical fluctuations $(\Delta_2^0\le 3)$  for nearly all the bins except the squeezed triangles where $3 < \Delta^0_{0} \le 5$.

The third and fourth rows of  Fig.~\ref{fig:mu_t_l=0&2}  show the quadrupole moment ($\ell=2$)  with $m=1$ and $m=2$, respectively. Unlike  $\bar{B}_{0}^0$  and $\bar{B}_{2}^0$, which are  positive throughout, we see that $\bar{B}_{2}^1$  and $\bar{B}_{2}^2$  have both positive and negative values. 
The magnitude  of  $\bar{B}_{2}^1$  and $\bar{B}_{2}^2$ are $\sim 4$ times  smaller than those of $\bar{B}_{0}^0$  and $\bar{B}_{2}^0$. Further, the patterns in the $\mu-t$ plane are also quite different. Considering $\bar{B}_2^1$, we see that 
it has small positive values for acute triangles ($\mu<t$) and negative values for obtuse triangles ($\mu>t$), with values close to zero for right-angled triangles $(\mu=t)$. Considering the right panel, we see that magnitude is the largest for squeezed triangles where $\bar{B}_{2}^1$ is negative. We also note that the values of $\bar{B}_{2}^1$ are in good agreement with those of   $[\bar{B}_2^1]_{\rm d}$.   The middle panel shows that $\Delta_2^1 \le 3$ for all the bins, with the exception of two bins (both at $\mu\approx 1$) where $3 < \Delta_2^1 \le 5$. Considering $\bar{B}_2^2$, we see that the results are very similar to those for $\bar{B}_2^1$. However, there is a difference that the values of $\bar{B}_2^2$ are positive over a large region of the $\mu-t$ plane as compared to $\bar{B}_2^1$.  We also see that $\Delta_2^2 \le 3$ for nearly all the bins except a few ($\sim 5$)  where $3 < \Delta_2^2 \le 5$.

Overall, we see that the estimated values $\bar{B}_{\ell}^m $ are in good agreement with the analytical predictions $[\bar{B}_{\ell}^{m}]_{\rm d}$ for the monopole moment $(\ell=0,m=0)$, and  all 
the quadrupole moments $(\ell=2,m=0,1,2)$. We have carried out the same analysis for all the higher multipole moments, namely  $(\ell=4,m=0,1,2,3,4)$ and $(\ell=6,m=0,1,2,3,4,5,6)$, that are predicted to be non-zero for our model \citep{2020_Bharadwaj}. The results for $\ell =4$ and $6$ are respectively presented in Figures  \ref{fig:mu_t_l=4} and \ref{fig:mu_t_l=6} of Appendix \ref{sec:app1}.  In all the cases,  we find that the values of $\bar{B}_{\ell}^m $ are in good agreement with the analytical predictions $[\bar{B}_{\ell}^{m}]_{\rm d}$. The values of $\Delta_{\ell}^{m}$ are mostly in the range  $\Delta_{\ell}^{m} \le 3$, and they are in the range $3 <\Delta_{\ell}^{m} \le 5$ for a few bins, indicating that the deviations between  $\bar{B}_{\ell}^m $ and $[\bar{B}_{\ell}^{m}]_{\rm d}$ are consistent with that  expected from statistical fluctuations.

% \begin{figure*}
%     \centering 
%     \includegraphics[width=1\textwidth]{fig/fft_ana_mut_l=2_12X12_pk2_k70.5_dBbysd_2.pdf} 
%     \caption{Estimator binning scheme }
%     \label{fig:mu_t_l=2}
% \end{figure*}

% \begin{figure*}
%     \centering 
%     \includegraphics[width=1\textwidth]{fig/zBS_with_mu_final_2.pdf} 
%     \caption{$k_1=2.1~\mpc^{-1}$ }
%     \label{fig:zbs_with_mu}
% \end{figure*}

\section{Summary and discussion}

The bispectrum is the lowest-order statistic sensitive to non-Gaussianity in the underlying density field. This is particularly relevant for the late-time matter distribution where non-Gaussianity develops due to non-linear evolution \citep{1980Peebles}, and the bispectrum contains a wealth of cosmological information \citep{Scoccimarro_1999,Feldman_2001,Gil_Marin_2014,Gagrani_2017,2022_Philcox_oneloop,ivanov_2023,Heinrich_2023}. The redshifted 21-cm signal from the Epoch of Reionization is also predicted to be highly non-Gaussian due to the emergence of large ionized regions  \citep{2005_Bharadwaj}, and the bispectrum \citep{Majumdar_2018,Hutter_2019,Watkinson_2021,2024_gill} is expected to carry a plethora of information regarding the nature of the ionizing sources.  It is, therefore, crucial to go beyond the power spectrum and consider the bispectrum and other higher-order statistics to extract the maximum information from observational data.

 %In real space, where distribution is statistically isotropic, it depends only on the shape and size of the triangle. However, t.. five parameters are required to define a triangle of a unique shape, size, and orientation relative to LoS
 The bispectrum is a function of the closed triangles formed by three $\kk$ vectors. Cosmological density fields are expected to be isotropic, and the bispectrum is predicted to depend only on the size and shape of the triangle, which we have parameterized using three parameters $k_1$ and $(\mu,t)$ respectively.  However, due to redshift space distortion, the observed bispectrum also depends on how the triangle is oriented with respect to the LoS, and we require five parameters to parameterize the redshift space bispectrum.  Here we have considered the redshift space bispectrum $B^s(\hat{\textbf{p}},k_1,\mu,t )$, where the unit vector  $\hat{\textbf{p}}$  \citep{2020_Bharadwaj} quantifies the orientation of the triangle with respect to the LoS. We have decomposed $B^s(\hat{\textbf{p}},k_1,\mu,t )$ into the spherical harmonic basis $Y_\ell^m(\pp)$ (Eq.~\ref{eq:bs_lm}),  and used the bispectrum multipole moments $\bar{B}_\ell^m(k_1,\mu,t )$ to quantify the RSD anisotropy. 
 
It is relatively straightforward to compute the bispectrum by looping over the available $\kk$ vectors, however, this is computationally very 
 expensive as the computing time scales as $\sim N_g^6$ for a 3D data cube of  $N_g$ grids in each dimension. This limitation is overcome by several  FFT-based fast bispectrum estimators where the computation time scales as $\sim N_g^3 \log{(N_g)}$ \cite{Sefusatti_thesis,Jeong_thesis,Scoccimarro_2015}. However, these estimators consider spherical shells in $\kk$ space, and they are restricted to the $m=0$ multipoles \cite{Federico_2023,ivanov_2023} of the redshift space bispectrum.  Considering the multipole moments with $\ell \le 6$, we only have four non-zero multipole moments if we are restricted to $m=0$, whereas we have a total of sixteen non-zero multipole moments if we also include the ones with $m \neq 0$ \citep{2020_Mazumdar}. The difficulty arises due to the $e^{im\phi}$ term in  $Y_\ell^m(\pp)$, as it is not possible to evaluate this with spherical shells in $\kk$ space. We have overcome this by dividing the $\kk$ space into $N_g$ planes perpendicular to the LoS and dividing each plane into $N_r$ annular rings. The combination of three rings and three planes contains all the triangles of a fixed size, shape, and orientation with respect to the LoS. We have used the FFT to evaluate the bispectrum for such a combination, and we average this over different orientations to evaluate the bispectrum multipole moments (Eq.~\ref{eq:bs_lm2}). The computation time for performing the FFT operations scale as $\mathcal{O}(N_g^3)$, while estimating all the bispectrum multipoles for all possible closed triangle configurations scales as $\mathcal{O}(N_g^4)$. Note that the estimator presented in this work is based on the plane-parallel approximation and assumes a fixed LoS. However, in realistic modern galaxy surveys that cover large angles of the sky, the LoS varies with the positions of galaxies, leading to the radial RSD. The effect due to the radial nature of the distortions may be important and can significantly influence the galaxy statistics. \citet{Yamamoto_2006} introduced a formalism to account for the varying LoS in the power spectrum multipoles, where the LoS is defined along the direction of one of the galaxies in a pair. Building on this, \citet{Scoccimarro_2015} proposed an estimator for the $m=0$ multipoles of the bispectrum that incorporates these radial distortions. We plan to incorporate it in our estimator in the future.

% The effect of the varying LoS may be important in the case of $m\neq 0$ multipoles, and
% In recent work,  \citet{ivanov_2023} analyzed the $m=0$ bispectrum multipoles for the BOSS survey, assuming a fixed LoS, stating that the impact of the LoS variations is negligible for typical survey volumes.
% \cite{ivanov_2023} studied the $m=0$ bispectrum multipoles for BOSS galaxy survey and ignored the effects due to varying LoS varying as the effect is expected to be small for typical survey sizes.
% A few previous studies have incorporated the varying LoS in 3PCF \citep{Garcia_2020}.
 
 We have validated our estimator using simulations. We have considered 50 realizations of a mildly non-Gaussian field that incorporates redshift space distortion. The analytical expressions for the bispectrum multipoles of this field are presented in \citep{2020_Bharadwaj}. We have used the estimator to calculate all the bispectrum multipoles that are predicted to be non-zero (up to $\ell=6,m=6$ in our case).  
 We find that the estimated values  $\bar{B}_\ell^m$ are in good agreement with the analytical predictions $[\bar{B}_\ell^m]_{\rm d}$ for all the cases.  The deviations of the estimates from the analytical predictions are within $\pm 3 \sigma$ for most of the bins,  consistent with their arising from statistical fluctuations. In a few bins, the deviations are in the range $\pm 3 \sigma$ to $\pm 5 \sigma$, which is also possibly due to statistical fluctuations. In some rare cases, mostly near the squeezed limit, we have deviations in the range 
 $\pm 5 \sigma$ to $\pm 7 \sigma$. However, in these cases, the fractional deviations are still of the order of $1 \%$ or smaller. 
 In conclusion, our estimator is able to calculate all the non-zero bispectrum multipole moments at a high level of accuracy. It is anticipated that including the $m \neq 0$ multipoles will add to the information already available from galaxy redshift surveys, and that is expected from future redshifted 21-cm observations.

\section*{Acknowledgements}
The authors thank the anonymous reviewer for the valuable suggestions and comments. We thank Abinash Kumar Shaw for the useful discussions. SSG acknowledges the support of the Prime Minister Research Fellowship (PMRF). We acknowledge the National Supercomputing Mission (NSM) for providing computing resources of ‘PARAM Shakti’ at IIT Kharagpur, which is implemented by C-DAC and supported by the Ministry of Electronics and Information Technology (MeitY) and Department of Science and Technology (DST), Government of India. 

\section*{Data availability}
The simulated data and the package involved in this work will be shared on reasonable request to the corresponding authors.

% The ongoing and upcoming surveys are expected to provide a huge volume (Gpc$^3$) of data. 

% The late-time growth of matter fluctuations is highly non-Gaussian due to non-linear gravitational instability. The non-Gaussian features are further imprinted on the distribution of tracers (galaxies, neutral hydrogen, Lyman-$\alpha$, etc) due to non-linear bias. Some inflationary theories also predict the non-Gaussianity in primordial fluctuations. It is crucial to extract this 

\label{sec:summary}

\bibliographystyle{apsrev4-2}
\bibliography{ref}
%--------------------------------------------------------------------
% \bsp
%\label{lastpage}

\appendix 

\section{Results for $\ell=4$ and $6$}
\label{sec:app1}

\begin{figure*}
    \centering 
    \includegraphics[width=1\textwidth]{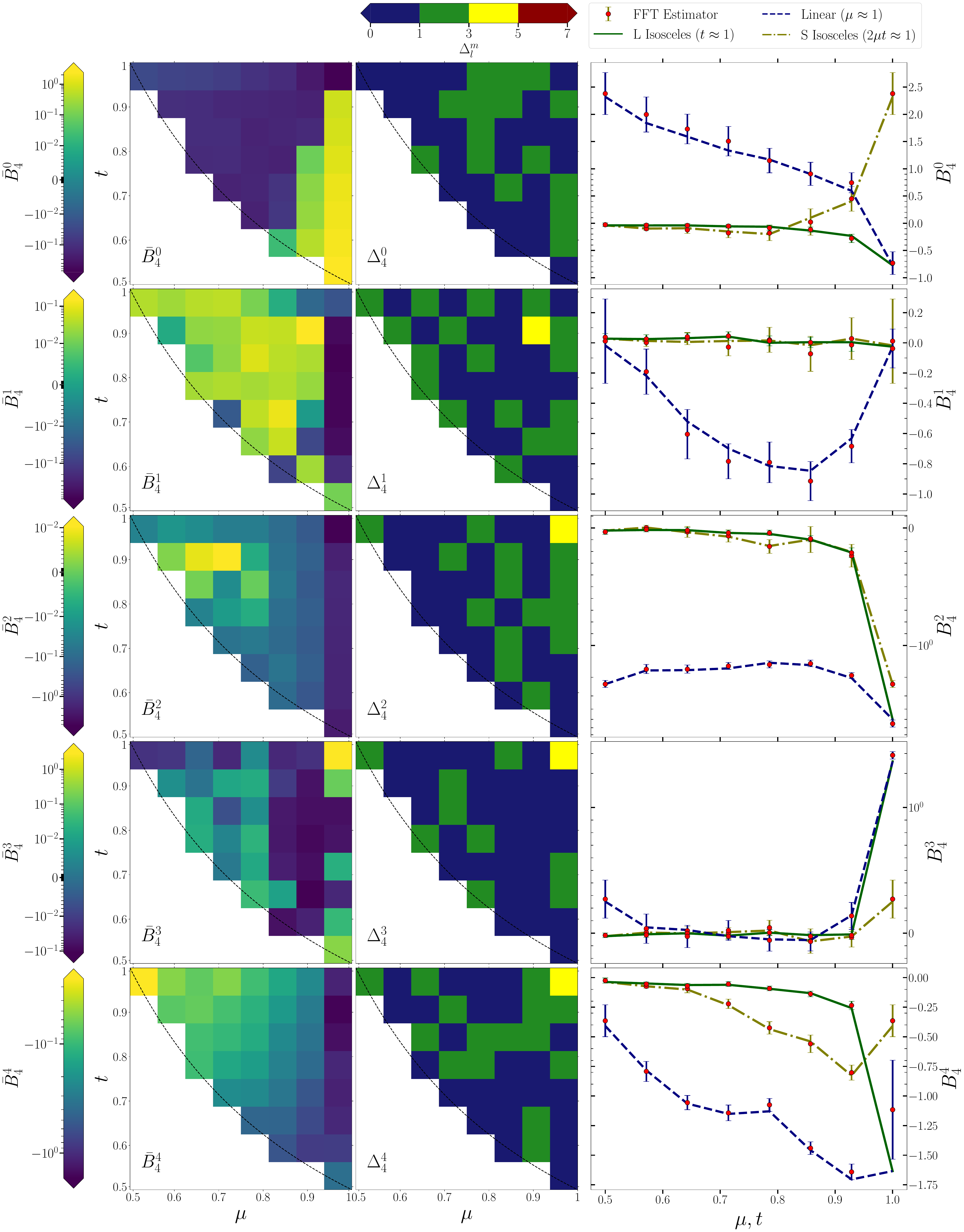} 
    \caption{ Results for $\ell=4$ multipoles (panels are same as in Fig.~\ref{fig:mu_t_l=0&2}). The error bars in the right panel are $3\sigma_{B_\ell^m}$ r.m.s fluctuations.}
    \label{fig:mu_t_l=4}
\end{figure*}

\begin{figure*}
    \centering 
    \includegraphics[width=.77\textwidth]{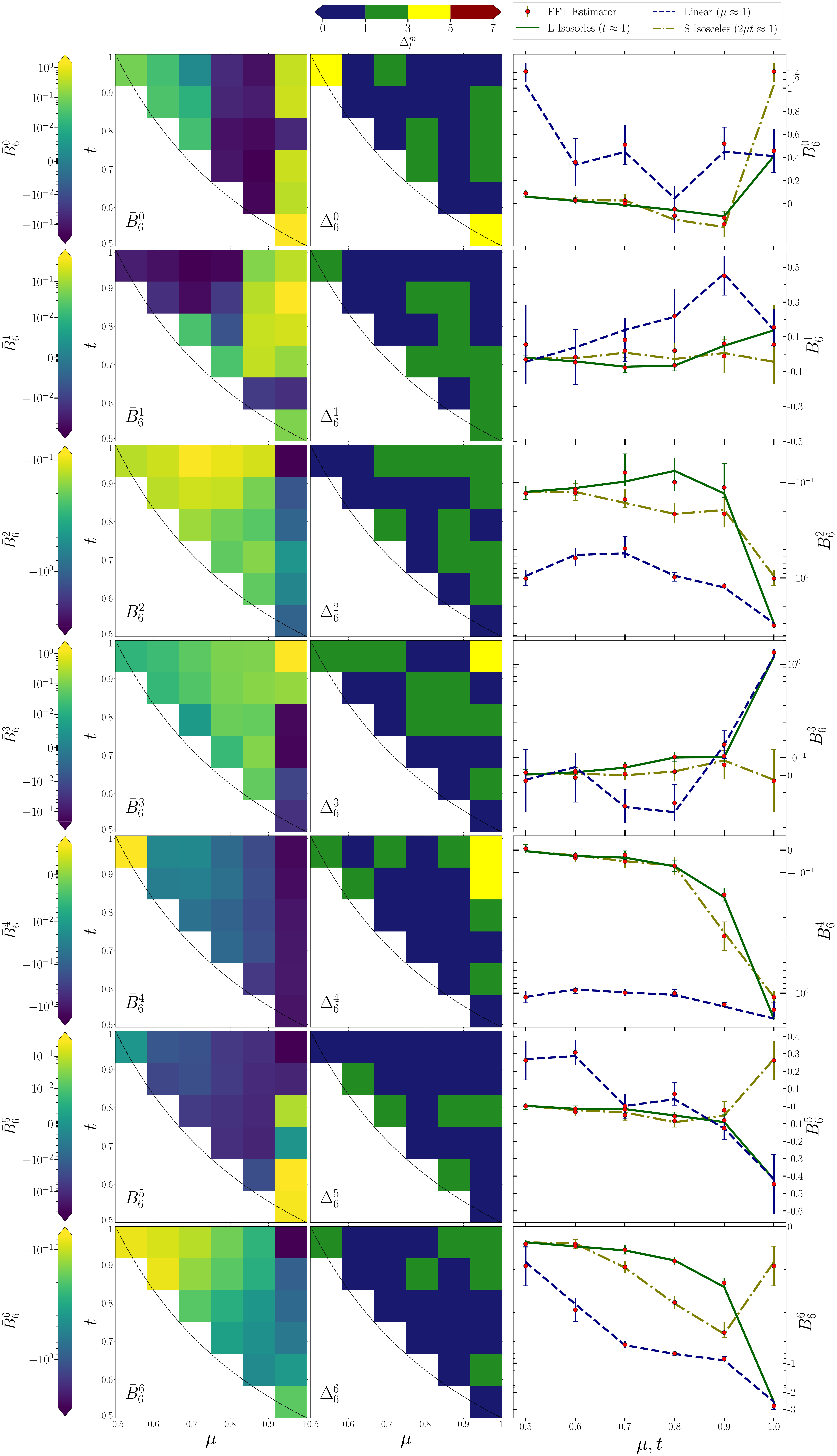} 
    \caption{Results for $\ell=6$ multipoles (panels are same as in Fig.~\ref{fig:mu_t_l=0&2}). The error bars in the right panel are $3\sigma_{B_\ell^m}$ r.m.s fluctuations.}
    \label{fig:mu_t_l=6}
\end{figure*}

Fig.~\ref{fig:mu_t_l=4} and \ref{fig:mu_t_l=6} shows the results for $\ell=4$ and $6$ bispectrum multipoles respectively. Note that these multipole moments have a smaller SNR as compared to the $\ell=0$ and $2$ multipoles, and 
we have divided the $\mu-t$ plane into wider bins to account for this. While the interval $0.5 \le \mu,t \le 1$ was divided into $12$ bins along each direction for both $\ell=0$ and $2$, we have used $8$ and $6$ bins for $\ell=4$ and $6$ respectively. For all the cases shown here, the estimates are in good statistical agreement with the analytical predictions. We find $\Delta_{\ell}^{m} \le 3$ for most of the cases, and it is in the range $3 <\Delta_{\ell}^{m} \le 5$ for a few bins near linear triangles. These deviations are roughly consistent with those expected from statistical fluctuations.

\end{document}